\documentclass[twocolumn]{aastex701}

\usepackage{amsmath}

\newcommand{\fpfsTESSobs}{\ensuremath{37\pm10\,\mathrm{ppm}}}
\newcommand{\fpfsHSTobs}{\ensuremath{69^{+14}_{-13}\,\mathrm{ppm}}}
\newcommand{\tbTESS}{\ensuremath{2566^{+94}_{-110}\,\mathrm{K}}}
\newcommand{\rTESS}{\ensuremath{0.996^{+0.039}_{-0.044}}}
\newcommand{\tbHST}{\ensuremath{2497^{+110}_{-119}\,\mathrm{K}}}
\newcommand{\rHST}{\ensuremath{0.970^{+0.044}_{-0.048}}}
\newcommand{\tbjoint}{\ensuremath{2524^{+77}_{-84}\,\mathrm{K}}}
\newcommand{\rjoint}{\ensuremath{0.980^{+0.032}_{-0.034}}}
\newcommand{\tbjointAgone}{\ensuremath{2407^{+91}_{-102}\,\mathrm{K}}}
\newcommand{\tpmax}{\ensuremath{2574^{+36}_{-35}\,\mathrm{K}}}
\newcommand{\tfullred}{\ensuremath{2015^{+28}_{-27}\,\mathrm{K}}}
\newcommand{\reflcoeff}{\ensuremath{105\pm6\,\mathrm{ppm}}}
\newcommand{\reflAgone}{\ensuremath{11\,\mathrm{ppm}}}
\newcommand{\flower}{\ensuremath{0.453}}
\newcommand{\ABupper}{\ensuremath{0.320}}
\newcommand{\fstarTESS}{\ensuremath{41.6\%}}
\newcommand{\fstarHST}{\ensuremath{21.3\%}}
\newcommand{\fstarTESSHST}{\ensuremath{62.9\%}}
\newcommand{\fstarTESSHSTfrac}{\ensuremath{0.629}}
\newcommand{\ABAgfactor}{\ensuremath{0.944}}
\newcommand{\fpfsTESSmax}{\ensuremath{38^{+4}_{-3}\,\mathrm{ppm}}}
\newcommand{\fpfsHSTmax}{\ensuremath{79\pm5\,\mathrm{ppm}}}
\newcommand{\fpfsMIRIsim}{\ensuremath{261\pm14\,\mathrm{ppm}}}
\newcommand{\miriPrecision}{\ensuremath{14\,\mathrm{ppm}}}

\accepted{September 25, 2026}

\begin{document}

\title{The First Hubble Detection of a Secondary Eclipse from a Rocky Exoplanet: the 5.4-hour planet TOI-2431 b}
\shorttitle{HST phase curve of TOI-2431 b}

\author[orcid=0000-0003-0562-6750,sname='Zieba']{Sebastian Zieba}
\altaffiliation{NASA Sagan Fellow}
\affiliation{Center for Astrophysics, Harvard \& Smithsonian, 60 Garden Street, Cambridge, MA 02138, USA}
\email[show]{sebastian.zieba@cfa.harvard.edu}  

\author[orcid=0000-0002-5494-3237]{Billy Edwards}
\affiliation{SRON, Space Research Organisation Netherlands, Niels Bohrweg 4, 2333 CA Leiden, The Netherlands}
\email{b.edwards@sron.nl}

\author[orcid=0000-0002-8964-8377]{Samuel N. Quinn}
\affiliation{Center for Astrophysics, Harvard \& Smithsonian, 60 Garden Street, Cambridge, MA 02138, USA}
\email{squinn@cfa.harvard.edu}

\author[orcid=0009-0002-9902-731X]{Leoni Janssen}
\affiliation{Leiden Observatory, Einsteinweg 55, 2333 CC Leiden}
\email{ljanssen@strw.leidenuniv.nl}

\author[orcid=0000-0001-6129-5699]{Nicolas B. Cowan}
\affiliation{Department of Earth \& Planetary Sciences, McGill University, 3450 rue University, Montréal, QC H3A 0E8, Canada}
\affiliation{Department of Physics, McGill University, 3600 rue University, Montréal, QC H3A 2T8, Canada}
\email{nicolas.cowan@mcgill.ca}

\author[orcid=0000-0003-0514-1147]{Laura Kreidberg}
\affiliation{Max Planck Institute for Astronomy, Königstuhl 17, 69117 Heidelberg, Germany}
\email{kreidberg@mpia.de}

\author[orcid=0000-0003-3204-8183]{Mercedes López-Morales}
\affiliation{Space Telescope Science Institute, 3700 San Martin Drive, Baltimore, MD 21218, USA}
\email{mlopez-morales@stsci.edu}

\author[orcid=0000-0003-4987-6591]{Lisa Dang}
\affiliation{Waterloo Centre for Astrophysics and Department of Physics and Astronomy, University of Waterloo, Waterloo, Ontario, Canada N2L 3G1}
\email{lisa.dang@uwaterloo.ca}

\author[orcid=0000-0002-6492-2085]{Luca Malavolta}
\affiliation{Dipartimento di Fisica e Astronomia “Galileo Galilei”, Università di Padova, Vicolo dell’Osservatorio 3, I-35122 Padova, Italy}
\affiliation{INAF - Osservatorio Astronomico di Padova, Vicolo dell'Osservatorio 5, IT-35122, Padova, Italy}
\email{luca.malavolta@unipd.it}

\author[orcid=0000-0002-0747-8862]{Yamila Miguel}
\affiliation{SRON, Space Research Organisation Netherlands, Niels Bohrweg 4, 2333 CA Leiden, The Netherlands}
\affiliation{Leiden Observatory, Leiden University, Leiden, The Netherlands}
\email{ymiguel@strw.leidenuniv.nl}

\author{T. Giang Nguyen}
\affiliation{Department of Physics, McGill University, 3600 rue University, Montréal, QC H3A 2T8, Canada}
\email{giang.nguyen@mail.mcgill.ca}

\author[orcid=0000-0001-7246-5438]{Andrew Vanderburg}
\affiliation{Center for Astrophysics, Harvard \& Smithsonian, 60 Garden Street, Cambridge, MA 02138, USA}
\email{avanderburg@cfa.harvard.edu}

\author[orcid=0000-0001-8749-1962]{Thomas G. Wilson}
\affiliation{Department of Physics, University of Warwick, Gibbet Hill Road, Coventry CV4 7AL, UK}
\email{thomas.g.wilson@warwick.ac.uk}

\author[orcid=0000-0002-8749-823X]{Mantas Zilinskas}
\affiliation{Jet Propulsion Laboratory, California Institute of Technology, Pasadena, CA 91109, USA}
\email{mantas.j.zilinskas@gmail.com}

\author[orcid=0000-0001-5848-6750]{Nicholas J. Connors} 
\affiliation{Waterloo Centre for Astrophysics and Department of Physics and Astronomy, University of Waterloo, Waterloo, Ontario, Canada, N2L 3G1}
\affiliation{Department of Physics and Trottier Institute for Research on Exoplanets, Universit\'{e} de Montr\'{e}al, Montreal, QC, Canada}
\email{nicholas.connors@umontreal.ca}

\author{Ian J.\ M.\ Crossfield}
\affiliation{Department of Physics and Astronomy, University of Kansas, Lawrence, KS, USA}
\email{ianc@ku.edu}

\author[orcid=0000-0002-1807-4441]{Oliver Herbort}
\affiliation{Department of Astrophysics, University of Vienna, Türkenschanzstr. 17, 1180 Vienna, Austria}
\email{oliver.herbort@univie.ac.at}

\author[orcid=0000-0001-9811-568X]{Adam L. Kraus}
\affiliation{University of Texas at Austin, Department of Astronomy, 2515 Speedway C1400, Austin, TX 78712, USA}
\email{alk@astro.as.utexas.edu}

\author[orcid=0000-0002-3286-7683]{Tim Lichtenberg}
\affiliation{Kapteyn Astronomical Institute, University of Groningen, Landleven 12, 9747 AD Groningen, The Netherlands}
\email{tim.lichtenberg@rug.nl}

\author[orcid=0009-0008-7799-7976]{Mariana Sastre} 
\affiliation{Kapteyn Astronomical Institute, University of Groningen, The Netherlands}
\email{m.c.villamil.sastre@rug.nl}  

\author[orcid=0000-0003-3191-2486]{Joost P. Wardenier}
\affiliation{Weltraumforschung und Planetologie, Physikalisches Institut, University of Bern, Gesellschaftsstrasse 6, 3012 Bern, Switzerland}
\email{joost.wardenier@unibe.ch}

\begin{abstract}

We present an HST/WFC3 G141 phase-curve observation of the ultra-short-period rocky planet TOI-2431 b and detect its secondary eclipse in the near infrared. Combined with the TESS optical eclipse, these measurements constrain the planet's dayside brightness temperature, reflected light contribution, albedo, and heat redistribution. 
We detect the eclipse in the optical and near infrared wavelengths of TESS and HST/WFC3 G141, with a planet-to-star flux ratio of \fpfsTESSobs\ and \fpfsHSTobs, respectively.
Assuming negligible reflected light, the joint TESS+HST fit yields a brightness temperature $T_{b,\rm joint} = \tbjoint$ and a brightness temperature ratio of $\mathcal{R}_{\rm joint}=T_{b,\rm joint}/T_{p,\max}=\rjoint$ consistent with the zero-albedo, no-redistribution limit $T_{p,\max} = \tpmax$. 
The HST phase curve is consistent with zero nightside emission and no measurable hotspot offset, consistent with inefficient day-night heat distribution.
Under this thermal-dominated interpretation, TOI-2431 b lies at the zero-albedo, no-redistribution limit and contrasts with previously observed lava worlds.
At optical and near-infrared wavelengths, however, thermal emission is degenerate with reflected light. 
For a geometric albedo of $A_g=0.1$, reflected light would contribute approximately \reflAgone\ in each bandpass, reducing the temperature to $T_{b,\rm joint}(A_g=0.1) = \tbjointAgone$.
Longer-wavelength observations are required to isolate the thermal component. 
Planned JWST/MIRI LRS observations are expected to substantially reduce this degeneracy and provide more direct constraints on heat redistribution and possible silicate-atmosphere spectral features.
This marks the first secondary eclipse detection of a rocky exoplanet using HST and establishes TOI-2431 b as a benchmark target for studying properties of lava worlds.
\end{abstract}


\section{Introduction} 

Lava worlds provide access to a regime of rocky planets with no analog in our solar system. We define lava worlds as planets with radii $R_p\lesssim1.9\,R_\oplus$ and substellar temperatures $T_{\rm irr}>1500$~K \citep{Coy2026}. These strongly irradiated planets are expected to be tidally locked, leading to permanently hot daysides that can sustain a permanent magma ocean. Evaporation from the molten surface can generate a rock-vapor, or silicate, atmosphere, in equilibrium with the underlying melt \citep{Schaefer2009, Leger2011, Miguel2011, Ito2015, Kite2016, Chao2021, Herbort2020, vanBuchem2023}. This creates a direct connection between the observable atmosphere, the molten surface, and the rocky interior.

Secondary eclipses and phase curves constrain the reflective and thermal properties of exoplanets. The eclipse measures the dayside planet-to-star flux ratio, while the phase variation constrains the day-night brightness contrast, nightside emission, and the longitudinal offset of the brightest region relative to the substellar point. These observables probe heat redistribution and atmospheric dynamics, although weak redistribution alone does not establish the absence of an atmosphere \citep[e.g.,][]{Parmentier2018}. 
At optical and near-infrared wavelengths, the measured dayside flux of the hottest rocky planets can include both reflected starlight and thermal emission \citep[e.g.,][]{LopezMorales2007,Malavolta2018,Mansfield2018}. 
Combining TESS or other optical data, like from Kepler or CHEOPS, with WFC3/G141 observations provides a useful wavelength combination: here TESS (0.6 -- 1.0 \micron) probes the short-wavelength side of the planetary thermal contribution, where reflected light can also significantly contribute, whereas the WFC3/G141 (1.1 -- 1.7 \micron) bandpass overlaps with the thermal emission peak of hot rocky exoplanets. For example, a 2500 K blackbody peaks at approximately 1.16 \micron. Similar combinations of optical and infrared observations have been used in the past to constrain the relative contributions of reflected light and thermal emission \citep[e.g.,][]{Mansfield2018, Zieba2022}.
At longer infrared wavelengths, thermal emission becomes dominant. Additionally, infrared emission spectroscopy can probe deviations from blackbody emission, caused by a rock-vapor, or silicate, atmosphere. In particular, models of silicate atmospheres predict prominent SiO and SiO$_2$ features near 9 and 7 \micron, respectively, which could constrain the composition of the underlying melt and should be accessible with JWST, in particular MIRI LRS \citep{Zilinskas2022, Zilinskas2023, vanBuchem2026}.

Existing observations indicate that lava worlds may differ from their colder, similarly sized counterparts. Whereas planets smaller than 1.9 $R_\oplus$ and with $T_{\rm irr} \leq$ 2000 K typically show temperatures and spectra consistent with low-albedo bare surfaces or relatively thin atmospheres ($<$10 bar) \citep[e.g.,][]{Kreidberg2025, Kreidberg2019, Crossfield2022, Greene2023, Xue2024, WeinerMansfield2024, Luque2025, Allen2025, Wachiraphan2025, Gillon2026, Zieba2026}, several lava worlds observed to date have dayside brightness temperatures significantly below their zero-albedo, no-redistribution limits, namely HD 3167 b \citep{Coy2026}, TOI-431 b \citep{Monaghan2025}, 55 Cancri e \citep{Hu2024}, K2-141 b \citep{Zieba2022}, and TOI-561 b \citep{Teske2025} (see Figure 4 for their irradiation temperatures). 
Infrared observations of 55 Cnc e with Spitzer and JWST indicate an atmosphere, in particular due to the low dayside temperature observed for the planet and its potentially variable dayside flux \citep{Demory2016, Angelo2017, Mercier2022, Hu2024, Patel2024}.  
Those low brightness temperatures indicate either effective heat transport, reflective surfaces or clouds, or a combination of these effects. 
Furthermore, optical observations tell us more about the reflective properties of these lava worlds. For example, the deep optical eclipse observed by Kepler for Kepler-10 b was attributed to an unusually reflective surface \citep{Batalha2011,Rouan2011}. 
Multiwavelength observations are needed to fully characterize these planets and distinguish between various processes.

TOI-2431 b is a favorable target for such follow-up observations. The planet orbits a nearby late-K dwarf every 5.4 hours, making it the shortest-period planet known with both measured radius and mass \citep{Tas2026}. It has a radius of $1.53 \pm 0.03$ $R_\oplus$, a mass of $6.2\pm1.6$ $M_\oplus$, and a bulk density of $9.4\pm2.5~\rm{g/cm}^3$. Its density is somewhat higher than expected for an exactly Earth-like composition but remains consistent, within the uncertainties, with a predominantly rocky planet \citep{Tas2026}. 
The system age is poorly constrained, \(2.0^{+9.1}_{-1.7}\) Gyr \citep{Tas2026}. If the system is in fact old, TOI-2431 b may have a more advanced stage of global interior solidification than younger lava worlds \citep{Sastre2026}.
The planet's irradiation or substellar temperature, $T_{\rm irr}$, is about 2849 K, implying that a substantial fraction of the planet's dayside is expected to be molten \citep{Leger2011, Chao2021, Zilinskas2022}. The apparent brightness of the host star \citep[$K_s$ = 7.55 mag;][]{Skrutskie2006} and the favorable planet-to-star contrast, quantified by an Emission Spectroscopy Metric \citep[ESM;][]{Kempton2018} of 27 \citep{Tas2026}, make TOI-2431 b one of the most favorable rocky planets for thermal-emission and phase curve characterization. 

In this paper, we present the analysis of an HST/WFC3 G141 phase curve of TOI-2431 b and combine it with optical photometry from TESS. We detect the planet's transit and secondary eclipse in both bandpasses and report the first secondary eclipse from a rocky exoplanet with HST. 
The TESS and HST observations together provide complementary constraints on the planet's dayside emission, reflected light contribution, and day-night contrast.
A planned full-orbit JWST/MIRI LRS phase curve as part of GO 8864 \citep[PI: L. Dang,][]{Dang2025} will extend the wavelength coverage to the mid-infrared, where the emission of the planet is dominated by the thermal component over the reflected one.

\section{Observations and Data Analysis}

\subsection{HST Observations}

We observed a phase curve of TOI-2431 b with the Wide Field Camera 3 (WFC3) aboard the Hubble Space Telescope (HST) using the G141 grism, which covers approximately 1.1 -- 1.7 \micron. The observations were taken on February 4, 2022 as part of General Observer (GO) Program 16660 \citep[PI: S. Quinn;][]{Quinn2021} and spanned 11 HST orbits. We used round-trip spatial scan mode, alternating between forward and reverse scan directions \citep{McCullough2012, Deming2013, Wakeford2013}. Spatial scanning distributes the observed light across a larger detector area, allowing for longer integrations while avoiding saturation and therefore improving the observing efficiency. The two scan directions exhibit a direction-dependent offset, which we account for in our light curve fitting model. 

The observations were taken with the 512x512 subarray (\texttt{SQ512SUB}) and had a scan rate of 0.223 arcsec/sec, corresponding to 1.7 pixels/sec. We used the SPARS25 read-out mode with NSAMP = 7 (eight reads, including the initial zeroth read), yielding an integration time of 138.38 seconds for the full up-the-ramp integration. Each HST orbit began with an undispersed direct image of the star taken with the WFC3 F139M narrow-band filter. These images were taken to locate the spectral trace. The direct image was then followed by 13 G141 spectroscopic exposures per HST orbit.  

\subsection{TESS Observations}
TOI-2431 was observed by TESS \citep{Ricker2015} in Sectors 31, 42, 43, 70, and 71, spanning October 2020 to November 2023. We use the SPOC \citep{Jenkins2016} light curves obtained at 120 seconds cadence in Sectors 31, 42, and 43 and at 20 seconds cadence in Sectors 70 and 71. The data were retrieved from MAST using the \texttt{lightkurve} package \citep{Lightkurve2018}.

\subsection{\texttt{PACMAN} HST Reduction}
\label{sec:pacman}

We downloaded the calibrated WFC3 IR Intermediate MultiAccum (IMA) files from the Mikulski Archive for Space Telescopes (MAST)\footnote{\url{https://mast.stsci.edu/search/hst}}. The files were processed with version 3.7.3 of the \texttt{calwf3} pipeline, which applies calibrations including dark subtraction, linearity correction, and flat fielding to each individual readout \citep{Pagul2024}. We reduced the IMA files with the open-source pipeline \texttt{PACMAN} \citep{Zieba2022_PACMAN}, following the reduction methods developed in previous WFC3 spatial-scan studies \citep[e.g.,][]{Kreidberg2014GJ1214, Kreidberg2014W43, Kreidberg2015W12, Kreidberg2018W103b}. \texttt{PACMAN} has also been used in several recent WFC3 analyses \citep[see e.g.,][]{Bachmann2025,Kahle2025}.

We take the difference between consecutive non-destructive reads and treat each difference image as an independent subexposure. We mask pixels for which data-quality bits 4 and 512 are set \citep{Pagul2024}, subtract the background, and optimally extract each difference image following \citet{Horne1986}. We use an extraction aperture extending 21 pixels beyond the upper and lower edges of the spectrum and sum the extracted difference spectra to obtain the final 1D spectrum for each exposure.

\begin{figure*}
    \centering
    \includegraphics[width=1\linewidth]{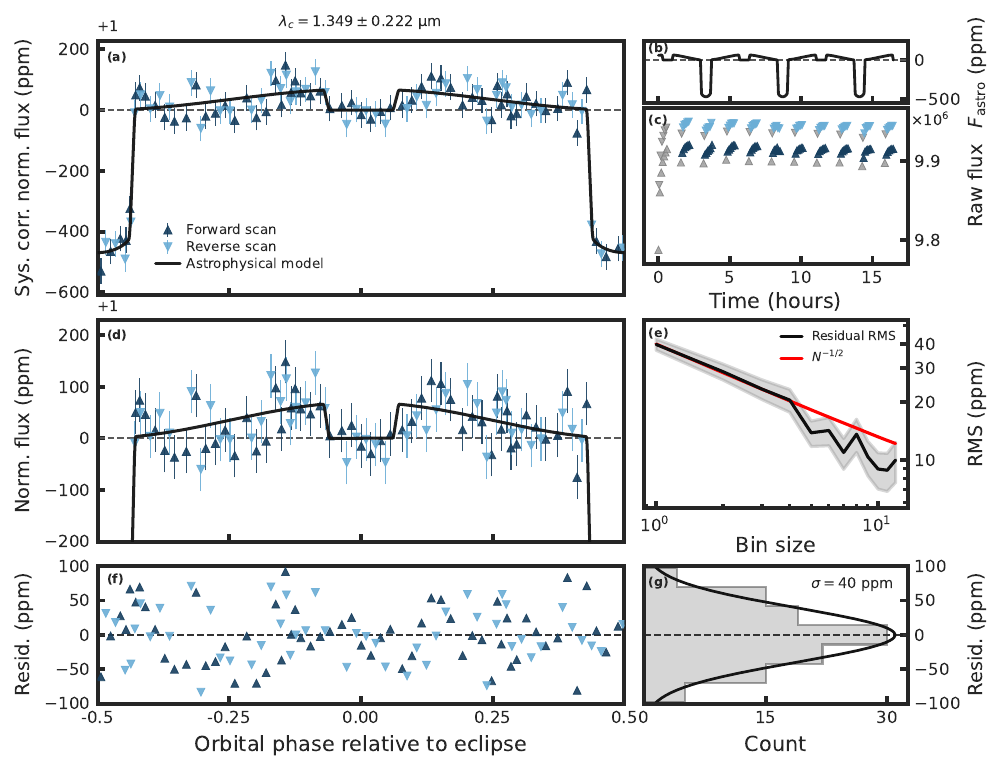}
    \caption{HST/WFC3 G141 white-light phase curve of TOI-2431 b, covering 1.127--1.570 \micron. For all subplots, we show the unbinned integrations with their 1$\sigma$ error bars that have been rescaled with a multiplicative factor to account for under- or overestimation of the pipeline's integration uncertainties. Note that panels (a), (d), and (f) are phase-folded. (a) Systematics-corrected, normalized, and phase-folded light curve, with the secondary eclipse centered at phase zero. Dark-blue upward triangles and light-blue downward triangles denote forward and reverse spatial scans, respectively, and the black curve shows the best-fitting astrophysical model. (b) The best-fitting astrophysical model, $F_{\rm astro}$, as a function of time since the first observed integration. (c) Uncorrected extracted \texttt{PACMAN} white-light flux. Grey points indicate observations that were removed in the fitting stage following previous work, because they are either in the first orbit or the first forward or reverse scan in an orbit. (d) Zoom in on panel (a), showing the phase variation and secondary eclipse. In panels (a) and (d), the horizontal dashed line denotes the stellar baseline flux. (e) RMS of the residuals as a function of temporal bin size. The black curve shows the measured RMS, the gray region its uncertainty, and the red curve the $N^{-1/2}$ expectation for uncorrelated noise, normalized to the unbinned residual RMS. (f) Residuals from the best-fitting light-curve model. (g) Distribution of the residuals; the black curve shows a Gaussian with $\sigma$=40 ppm, which corresponds to the standard deviation of the best-fitting residuals.}
    \label{fig:hst_lc}
\end{figure*}

We wavelength-calibrate the spectra by cross-correlating them with the G141 throughput multiplied by a smoothed \texttt{PHOENIX} stellar spectrum \citep{Allard2012}, selected to match the stellar parameters of \citet{Tas2026} and obtained using \texttt{pysynphot} \citep{pysynphot2013}. Because part of the spectral trace fell outside the detector, we restrict the broadband light curve to 1.127--1.570 \micron. Additional tests used to select the background, extraction aperture, and wavelength range are described in Appendix \ref{app:sec1_reduction}. The resulting light curve is shown in panel (c) of Figure \ref{fig:hst_lc}.

Our full fitting model can be described by
\begin{equation}
    F(t) = F_\textrm{transit}(t) F_\textrm{emission}(t) F_\textrm{sys,o}(t) F_\textrm{sys,v}(t),
\end{equation}
where \(F_{\mathrm{transit}}(t)\) is a transit model calculated with \texttt{batman} \citep{Kreidberg2015}, \(F_{\mathrm{emission}}(t)\) describes the secondary eclipse and phase variation, and \(F_{\mathrm{sys,o}}(t)\) and \(F_{\mathrm{sys,v}}(t)\) describe the orbit-long and visit-long systematics, respectively.
The emission consists of a \texttt{batman} \citep{Kreidberg2015} eclipse model, that is normalized to 1 during eclipse, together with cosine and sine phase curve terms:

\begin{equation} \label{eqn:fitting_model_emission}
    F_\textrm{emission}(t) = 1 + (F_\textrm{eclipse}(t) - 1) (1 + F_\textrm{pc}(t)),\\
\end{equation}

\begin{equation}
\begin{aligned}
\label{eq:Fpc}
F_{\mathrm{pc}}(t)
={}&
\frac{\mathrm{AmpCos}}{2}
\left[
\cos\left(
\frac{2\pi (t-t_{\mathrm{sec}})}{P_{\mathrm{orb}}}
\right)
-1
\right]
\\
&+
\frac{\mathrm{AmpSin}}{2}
\sin\left(
\frac{2\pi (t-t_{\mathrm{sec}})}{P_{\mathrm{orb}}}
\right).
\end{aligned}
\end{equation}

This parameterization ensures that \((\mathrm{AmpCos},\mathrm{AmpSin}) = (1,0)\) corresponds to zero nightside flux and no phase offset. When both coefficients are allowed to vary, we define the phase curve amplitude as \(\mathrm{Amp} = \sqrt{\textrm{AmpCos}^2+\textrm{AmpSin}^2}\) and the offset as \(\phi = \operatorname{atan2}(\textrm{AmpSin}, \textrm{AmpCos}) \). Therefore, \(\mathrm{Amp}=1\) corresponds to a full peak-to-trough phase-curve amplitude equal to the eclipse depth. As part of our model comparison, we allow both coefficients to vary, thereby testing for nightside emission and a longitudinal phase-curve offset rather than fixing either quantity a priori.

It has been shown that the choice of the visit-long systematic trend can impact the recovered transit or eclipse depth \citep[e.g.,][]{Stevenson2014,Guo2020,Edwards2024}. We therefore tested linear, quadratic, exponential, and logarithmic visit-long trends. We additionally model the orbit-long WFC3 ramp with an exponential function and allow the first fitted orbit to have an additional ramp term; the full functional forms of these equations and justifications are given in Appendix \ref{app:sec1_1} and \ref{app:sec1_2}.

We discard the very first HST orbit of the observation and also the first two exposures (one in each scan direction) of each HST orbit because they show stronger systematics, both being common practice \citep[see e.g.,][]{Deming2013, Kreidberg2014GJ1214, Wakeford2017}. After this, the remaining data cover all orbital phases, mapped over approximately three full orbits of the planet (see panel (a) of Figure \ref{fig:hst_lc}).

The \texttt{batman} transit and eclipse models need the following parameters: orbital period $P_{\rm orb}$, transit midpoint time $t_0$, planet-to-star radius ratio $R_p/R_s$, ratio of semi-major axis to stellar radius $a/R_s$, orbital inclination $i$, linear limb darkening $u_1$ (or quadratic limb darkening $u_1$ and $u_2$), and eclipse depth $f_p/f_s$. We also account for the light travel time across the orbit, which shifts the eclipse timing by approximately 6 seconds. For our fits, we also use normal priors on $P_{\rm orb}$, $R_p/R_s$, $a/R_s$, $i$, and $R_s$ based on the literature values reported in \citet{Tas2026}. The stellar radius ($R_s$) is needed to calculate the correction of light travel time. Finally, we also add a multiplicative factor to the uncertainties of our integrations, $\sigma_\textrm{multi}$, which allows the integration uncertainties to be rescaled during the fit.

We estimate the model parameters and their uncertainties using the dynamic nested sampling implementation in the open-source code \texttt{dynesty} \citep{Skilling2004, Skilling2006, Higson2019, Speagle2020, Koposov2025}. We also use the Bayesian evidence, $\log \mathcal{Z}$, returned by the nested sampler to guide our model selection.

After we perform the model comparisons listed in Appendix \ref{app:sec1}, we adopt the preferred systematic and astrophysical model for our final HST fit.
For the astrophysical parameters, we fit for $t_0$, $P_{\rm orb}$, $R_p/R_s$, $a/R_s$, $i$, $f_p/f_s$, $u_1$, and $R_s$.
The visit- and orbit-long systematic parameters are listed in Appendix \ref{app:sec1}.
This final fit finds a WFC3 eclipse depth of \(f_p/f_s = \fpfsHSTobs\). 
We perform a robustness test where we fix $R_p/R_s$ to the literature value instead of fitting it simultaneously with $f_p/f_s$. We find essentially unchanged results, with negligible impact on the measured eclipse depth (i.e., shift of less than 1 ppm and a decrease in uncertainty of 7\%).
The unbinned residuals of the best-fitting light curve have an RMS of 40 ppm, corresponding to 1.21 times the predicted photon-noise limit of 33 ppm. 
The residual RMS decreases approximately as \(N^{-1/2}\) over the tested bin sizes, with no clear evidence for time-correlated noise (see Fig. \ref{fig:hst_lc}).
In an alternative fit in which the phase-curve amplitude and offset are allowed to vary, we find \(\mathrm{Amp}=\mathrm{0.88^{+0.24}_{-0.19}}\) and \(\phi=\mathrm{4.1{^{\circ}}^{+12.3{^{\circ}}}_{-12.6{^{\circ}}}}\), consistent with the zero-nightside, no hotspot offset solution. We also find an orbit consistent with circular, with a \(3\sigma\) upper limit of \(e<0.19\). 

\subsection{\texttt{Iraclis} HST Reduction}

We performed a second, independent reduction with \texttt{Iraclis} \citep{Tsiaras2016, Tsiaras2018}. The methodology of the pipeline is described in detail in previous works and, like \texttt{PACMAN}, this open-source code has been widely used, including to produce the largest population studies of exoplanet atmospheres to date \citep{Changeat2022, Edwards2023}. Crucially, \texttt{Iraclis} is distinct from \texttt{PACMAN} as it starts with the raw files and performs its own calibration (dark subtraction, linearity correction, etc.) on them. In this reduction, we also split the ``up-the-ramp samples'' to get the difference between each read and, for consistency with the \texttt{PACMAN} reduction, we also extract the white light curve across a wavelength range of 1.127 -- 1.570 \micron. 

Subsequently, we fit this light curve using a custom-built code. The transits and eclipses are modeled using \texttt{pylightcurve} \citep{pylightcurve} while the phase curve is also represented by a cosine (i.e., Equation \ref{eq:Fpc}). We fix the limb-darkening coefficients and use the Claret 4-coefficient model \citep{claret_ldc} which we calculated using \texttt{ExoTETHyS} \citep{morello_exotethys}. Instead of fitting for $a/R_s$, here we fit the stellar radius ($R_s$) and mass ($M_s$), using Kepler's third law to determine $a$. These values, along with the other system parameters ($P_{\rm orb}$, $i$), are taken from \citet{Tas2026} and Gaussian priors were used. The systematics models are identical, with the exception of the orbit-long trend (Equation \ref{eqn:fitting_model_orbit}). Instead of fitting for $D(t)$, we fit one set of ramp parameters for the first orbit ($r1_{for}$ and $r2_{for}$) and another set for the subsequent orbits ($r1_{oor}$ and $r2_{oor}$). To explore the parameter space, we use \texttt{pymultinest} \citep{feroz_multinest,buchner_multinest}. 

In this case, we do not explore the impact of different fitting assumptions (e.g., the choice of long-term trend), but instead provide one fit as a direct comparison. Hence, we use a linear long-term trend, assume that the orbit is circular, and assume that there is no nightside emission or hotspot offset. From this fit, we achieve a WFC3 eclipse depth of 
$60\pm14$ ppm, consistent with the \texttt{PACMAN} result. We adopt the \texttt{PACMAN} result as the fiducial value because the full model-selection analysis was performed for that reduction.
For more information on the \texttt{Iraclis} reduction or the phase-folded \texttt{Iraclis} light curve, see Appendix Section \ref{app:sec2} with Figure \ref{fig:irac}.

\subsection{TESS Reduction}
We work with the TESS PDCSAP light curves and keep only data points flagged with $\mathrm{QUALITY}=0$. We then clean the light curves by removing $5\sigma$ outliers using iterative sigma clipping. We also exclude transits that are cut off at sector boundaries. We then jointly fit all five sectors with a custom-written fitting framework, independent from the HST analysis, by fitting for the \texttt{batman} transit and eclipse models and a cosine phase-curve term. We use \texttt{dynesty} to estimate the posteriors of the fitted parameters.

We account for time-correlated variability in each TESS sector using a Gaussian process with a Matérn-3/2 kernel implemented with \texttt{celerite2} \citep{ForemanMackey2017}. The astrophysical parameters ($P_{\rm orb}, t_0, R_p/R_s, a/R_s, i, f_p/f_s$) are shared between all sectors with broad uniform priors. Each sector has an independent flux normalization constant, GP amplitude, GP correlation timescale, and an integration uncertainty rescaler. We also supersample Sectors 31, 42, and 43, which had only 120-second cadence data, by a factor of six. We do not supersample Sectors 70 and 71, which have the higher 20-second cadence data available.

Allowing $\mathrm{AmpCos}$ to vary with broad uniform priors, and therefore allowing for nightside emission, results in $\mathrm{AmpCos} = 1.12^{+0.44}_{-0.31}$, consistent with unity. The corresponding eclipse depth is 
$33\pm11$ ppm, which is consistent with the result when fixing $\mathrm{AmpCos} = 1$. This fixed model is also weakly preferred by $\Delta\log\mathcal{Z}=1.3$. We therefore, due to the simpler model, adopt $\mathrm{AmpCos}=1$, corresponding to zero nightside emission and a phase-curve amplitude equal to the eclipse depth. Our final fit yields a TESS eclipse depth of $f_p/f_s =$ \fpfsTESSobs. In Figure \ref{fig:TESS} we show the phase-folded TESS phase curve.

\section{Interpretation}

\subsection{Thermal interpretation of the dayside and phase curve}

In order to put our measured planet-to-star flux ratio $f_p/f_s$ into context, we first convert it into a band-integrated dayside brightness temperature, $T_{b}$. We calculate the expected planet-to-star flux ratio as

\begin{equation}
\label{eq:fpfs}
\frac{f_p}{f_{s}}
=
\left(\frac{R_p}{R_{s}}\right)^2
\frac{
\displaystyle
\int_{\lambda_{\min}}^{\lambda_{\max}}
\pi B_{\lambda}(T_{b})\,
\frac{\lambda}{hc}\,
W_{\lambda}\,
\mathrm{d}\lambda
}{
\displaystyle
\int_{\lambda_{\min}}^{\lambda_{\max}}
F_{\lambda,\star}
\left(
T_{\star},
\log g,
[\mathrm{M/H}]
\right)\,
\frac{\lambda}{hc}\,
W_{\lambda}\,
\mathrm{d}\lambda
}.
\end{equation}

Note that for the calculations in this subsection, we initially assume that both eclipse depths are entirely thermal.

Here, $B_{\lambda}(T_{b})$ is the Planck function with planetary brightness temperature $T_b$, and $W_{\lambda}$ is the wavelength-dependent instrument throughput, which we access using \texttt{stsynphot} \citep{stsynphot2020}. Applying Equation \ref{eq:fpfs} to the TESS and HST/WFC3 G141 eclipse depths yields $T_{b, \rm TESS}$ and $T_{b, \rm HST}$, respectively. $F_{\lambda,\star}(T_{\star},\log g,[\mathrm{M/H}])$ represents the stellar model, for which we used the PHOENIX NewEra model grid \citep{Hauschildt2025}, accessed and interpolated with \texttt{speclib} \citep{Rackham2024, Rackham2026}. Within \texttt{speclib}, we used the \texttt{newera\_lowres} product for the TESS  and HST/WFC3 calculations, which covers 0.25 -- 2.5 \micron\ and therefore encompasses both bandpasses. For predictions at JWST/MIRI wavelengths, we use the \texttt{newera\_jwst} product, which extends from 0.6 to 28.5 \micron. We adopt the literature values and uncertainties for the stellar temperature $T_{\star}$, the surface gravity $\log g$, and the metallicity $[\mathrm{Fe/H}]$ (we use $[\mathrm{Fe/H}]$ as a proxy for $[\mathrm{M/H}]$, which is needed for generating the stellar models) from \citet{Tas2026}. 

In order to calculate the band-integrated dayside brightness temperature, we first draw the eclipse depth, $R_p/R_s, T_{\star},\log g,$ and $[\mathrm{M/H}]$ from their posteriors or their literature values. For each draw, we calculate the band-integrated stellar flux for the respective instrument from the interpolated NewEra spectrum and essentially numerically calculate $T_b$ from Equation \ref{eq:fpfs}. By using the WFC3/G141 bandpass and the wavelengths of 1.127--1.570 \micron\ we calculate a band-integrated dayside brightness temperature of $T_{b,\rm HST} = \tbHST$.

A planet that is in equilibrium between the energy it receives from the host star and the radiation it reradiates will have the temperature:
\begin{equation}
\label{Tday_def}
T_{p, day}
=
T_{\star}
\left(\frac{a}{R_s}\right)^{-1/2}
f^{1/4}
\left(1-A_B\right)^{1/4},
\end{equation}
where $A_B$ is the planet's Bond albedo, and the heat redistribution factor is given by $f$. In this convention, $f=1/4$ corresponds to uniform redistribution over the planet, whereas $f=2/3$ corresponds to no heat redistribution and instantaneous reradiation \citep[e.g.,][]{LopezMorales2007, Seager2010}. 
The zero-albedo, no-redistribution dayside limit is therefore
\begin{equation}
\label{Tmax_def}
T_{p,\max}
=
T_{\star}
\left(\frac{a}{R_{s}}\right)^{-1/2}
\left(\frac{2}{3}\right)^{1/4}.
\end{equation}

We also define the irradiation temperature as \citep{Cowan2011}
\begin{equation}
\label{Tirr_def}
T_{\rm irr}
=
T_{\star}
\left(\frac{a}{R_{s}}\right)^{-1/2}.
\end{equation}

For TOI-2431 b, we calculate $T_{p,\max} = \tpmax$. We can then define the brightness temperature ratio given by
\begin{equation}
\mathcal{R}_{\rm X}
=
\frac{T_{b,\rm X}}{T_{p,\max}}, X \in \{\rm TESS, \rm HST, \rm joint\}.
\end{equation}

We can calculate this brightness temperature ratio using the same procedure as earlier, where we repeatedly draw all parameters from Equation \ref{eq:fpfs} to determine $T_b$, but additionally sample $a/R_s$ to calculate $T_{p,\max}$. With that we derive $\mathcal{R}_{\rm HST} = \rHST$ which is consistent with the zero-albedo, no-redistribution limit, i.e., $\mathcal{R} = 1$.

When performing the analogous calculation for the TESS bandpass, we obtain $T_{b,\rm TESS} = \tbTESS$ and therefore $\mathcal{R}_{\rm TESS} = \rTESS$.

In order to determine the joint thermal-only constraints, we fit a single blackbody to both eclipse depths simultaneously using the MCMC sampler \texttt{emcee} \citep{ForemanMackey2013}. We again account for the uncertainties on the eclipse depths, $R_p/R_s, T_{\star},\log g,$ and $ [\mathrm{M/H}]$. We use 64 walkers for 40,000 steps, discard the first 10,000 steps, and thin the chains by a factor of five. From that we calculate $T_{b,\rm joint} = \tbjoint$ and therefore $\mathcal{R}_{\rm joint} = \rjoint$.

Under the thermal-only, single-temperature blackbody case, the joint eclipse measurements constrain the combination $f\,(1-A_B)$ (see Equation \ref{Tday_def}). We obtain $f > \flower$ (2$\sigma$) and $A_B < \ABupper$ (2$\sigma$), with the posteriors peaking at $f =2/3 $ and $A_B = 0$, respectively, which are also the physical bounds of those parameters ($1/4 \leq f \leq 2/3$ and $0 \leq A_B \leq 1$). See Appendix Figure \ref{fig:miri_degeneracy} panel (a) for these posteriors.

Taken together, these results, when assuming that the TESS and HST bandpasses are dominated by thermal emission, paint a consistent picture: The joint TESS+HST brightness temperature lies at the zero-albedo, no-redistribution limit, and the energy budget posteriors similarly also favor high $f$ and low $A_B$. Furthermore, the HST phase curve is consistent with zero nightside and no hotspot offset. This interpretation assumes that the measured eclipse depths are thermal-dominated. In Section \ref{sec:degeneracy}, we explore the degeneracy with reflected light in these wavelengths.

\begin{figure}
    \centering
    \includegraphics[width=1\linewidth]{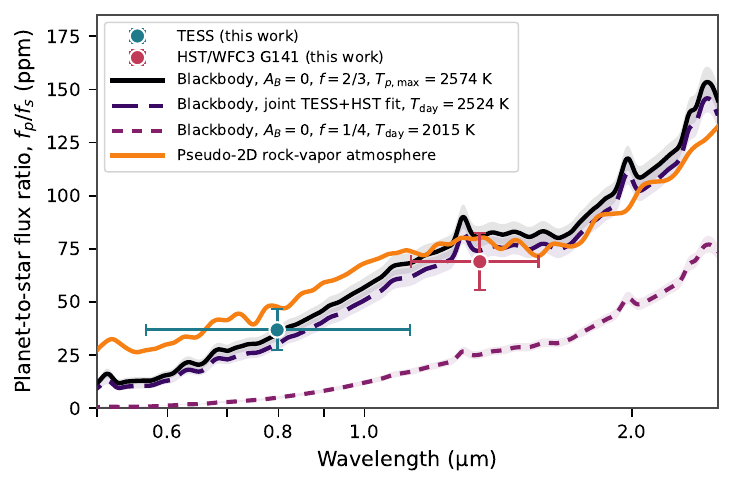}
    \caption{Measured planet-to-star flux ratios of TOI-2431 b in the TESS and HST/WFC3 G141 bandpasses compared with blackbody planetary emission models. The blue and red circles show the measured TESS and HST eclipse depths, respectively. The solid black curve shows the zero Bond albedo, no-redistribution model \(f=2/3\) and \(T_{p,\max}=\tpmax\). The long-dashed curve shows the joint TESS+HST thermal-only blackbody solution, with \(T_{\rm day}=\tbjoint\). The short-dashed curve shows the zero-albedo, full-redistribution limit \(f=1/4\) and \(T_{\rm day}=\tfullred\). The shaded region around these models shows the uncertainty in this prediction, propagated from the stellar and orbital parameters. The orange curve shows the disk-integrated pseudo-2D rock-vapor atmosphere forward model described in Section \ref{pseudo2Dmodeling}. For the blackbody cases, the planet is treated as a blackbody. Any spectral features in these predicted planet-to-star flux ratios therefore arise from the stellar spectrum.}
    \label{fig:spectrum}
\end{figure}

In Figure \ref{fig:spectrum}, we show the measured TESS and HST/WFC3 G141 planet-to-star flux ratios. We compare these measurements with model emission spectra in which the planet is approximated as a blackbody at different temperatures. Any spectral features seen in the modeled planet-to-star flux ratio spectra arise from the \texttt{PHOENIX NewEra} stellar model grid \citep{Hauschildt2025}, rather than from the planet, whose emission is assumed to be a blackbody. Under the assumption that both eclipse depths are entirely due to thermal emission, the measurements are consistent with a dayside temperature close to $T_{p,\max}$.

Assuming that $T_{p,\max}$ accurately describes the planet's dayside temperature, we predict thermal eclipse depths of approximately \fpfsTESSmax\ in the TESS bandpass and \fpfsHSTmax\ over the wavelength range used for the
HST/WFC3 G141 analysis. These predictions are consistent with the measured values of \fpfsTESSobs\ and \fpfsHSTobs, respectively. Finally, we also show a rock vapor model in Figure \ref{fig:spectrum}, which is currently consistent with the observations. We describe our setup for this model in Section \ref{pseudo2Dmodeling}.

\subsection{Degeneracy between thermal and reflected light}
\label{sec:degeneracy}

The reflected-light contribution to the planetary eclipse depth is given by
\begin{equation}
\begin{alignedat}{2}
\left(\frac{f_p}{f_s}\right)_{\mathrm{refl.}}
&=
A_g\left(\frac{R_p}{a}\right)^2
&\quad ={}& A_g
\left[
\frac{R_p/R_s}{a/R_s}
\right]^2
\\
&&\quad ={}& A_g \times \reflcoeff,
\end{alignedat}
\end{equation}
where $A_g$ is the geometric albedo in the observed bandpass \citep[e.g.,][]{LopezMorales2007}. At these short wavelengths, the measured eclipse depth is therefore degenerate between reflected light and thermal emission, i.e.,
\begin{equation}
\left(f_p/f_s\right)_{\mathrm{obs.}} = \left(f_p/f_s\right)_{\mathrm{refl.}}+\left(f_p/f_s\right)_{\mathrm{therm.}}.
\end{equation}

Under the wavelength-independent geometric albedo assumption ($A_{g,\mathrm{TESS}}=A_{g,\mathrm{HST}}=A_g$), the reflected-light contribution is identical in both bandpasses. It therefore cancels when the two eclipse depths are subtracted. The remaining measured difference between the HST and TESS eclipse depths is $32\pm17$ ppm. Blackbody temperatures of 2574 K ($A_B=0$, $f=2/3$) and 2015 K ($A_B=0$, $f=1/4$) predict differences of approximately 41 ppm and 19 ppm, respectively. A purely reflected-light signal would give zero difference. The TESS and HST observations are therefore able to reduce the thermal/reflected-light degeneracy, but the current observations are not precise enough to break it strongly. More precise observations or observations in other wavelength regimes will substantially reduce this degeneracy, as we discuss in Section \ref{sec:futureMIRI}.

\begin{figure}[h]
    \centering
    \includegraphics[width=1\linewidth]{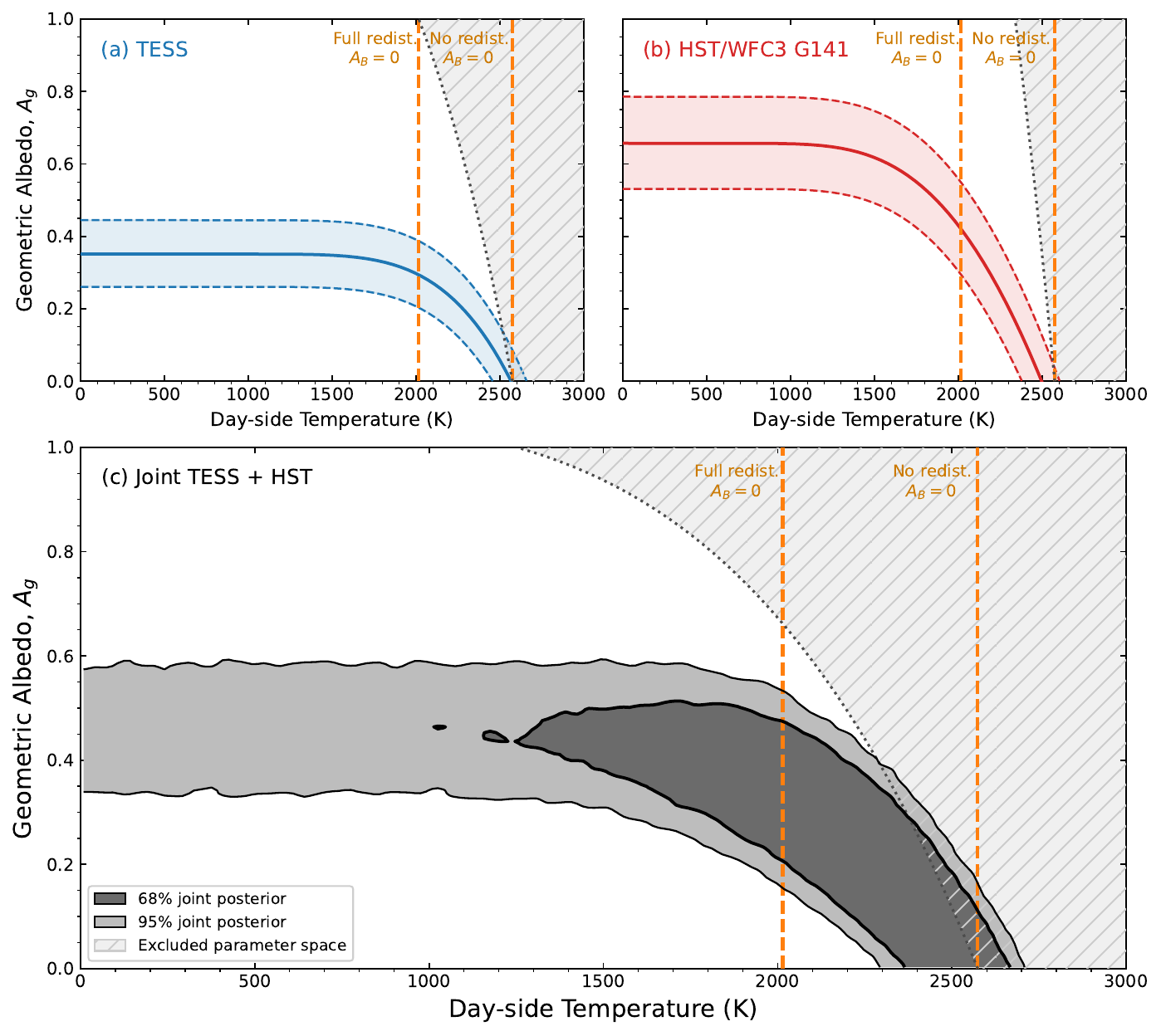}
    \caption{Constraints on the geometric albedo $A_g$ and the dayside temperature $T_{day}$ of TOI-2431 b from the measured eclipse depths. The top panels show the degeneracies from the TESS and HST measurements individually. The bottom panel shows the joint constraint. The two orange lines show the dayside temperature in case of no heat redistribution and zero albedo ($T_{day} = \tpmax$) and in case of full heat redistribution and zero albedo ($T_{day} = \tfullred$). The gray hatched region is inaccessible to stellar irradiation alone under the adopted Lambertian case. However, additional internal or tidal heating would permit temperatures within this excluded area \citep{Farhat2025}. 
    Similar graphical parameterizations have been used previously \citep[e.g.,][]{Keating2017,Malavolta2018,Singh2022}.}
    \label{fig:Ag_vs_Tday}
\end{figure}

In Figure \ref{fig:Ag_vs_Tday}, we show for our observed eclipse depths with TESS and HST, possible combinations of geometric albedo and dayside temperatures. As established in equation \ref{Tday_def}, the latter property is importantly a function of heat redistribution and the Bond albedo. Therefore, $T_{p, \rm{day}}$ (through $A_B$) and $A_g$ cannot be completely independent. Motivated by \cite{Schwartz2015}, we determine a lower bound on the Bond albedo for each assumed geometric albedo $A_g$ and with that determine the maximum dayside temperature caused by stellar irradiation. Using the \texttt{PHOENIX NewEra} models \citep{Hauschildt2025}, we calculate that we cover \fstarTESS\ and \fstarHST\ of the total stellar flux in the TESS and HST/WFC3 G141 bands. We then use \(A_s=qA_g\), where $A_s$ is the spherical albedo, $A_g$ the geometric albedo, and $q$ the phase integral. The Bond albedo is defined as:
\begin{equation}
A_B
=
\frac{
\int A_s(\lambda)F_{\lambda,\star}\,\mathrm{d}\lambda
}{
\int F_{\lambda,\star}\,\mathrm{d}\lambda
}.
\end{equation}
The minimum Bond albedo is therefore \(A_{B,\min}=A_s F_\star=qA_gF_\star\), where this extreme case assumes $A_s=0$ outside of the observed wavelength range. Assuming a single, wavelength-independent geometric albedo across the wavelengths (\(A_{g,\mathrm{TESS}}=A_{g,\mathrm{HST}}=A_g\)) we get \(F_{\star,\mathrm{TESS+HST}}=\fstarTESSHSTfrac\), therefore meaning that the combined TESS and adopted WFC3 wavelength ranges contain approximately \fstarTESSHST\ of the stellar bolometric flux. With this we get \(A_B\geq\fstarTESSHSTfrac\,qA_g\) and assuming a Lambertian, isotropically diffuse reflector \citep[e.g.,][]{Seager2010} for the planet (\(q=3/2\)) it follows \(A_B\geq\ABAgfactor\,A_g\) and therefore 
\begin{equation}
T_{\mathrm{day},\max}
=
T_{p,\max}
\left(
1-A_B
\right)^{1/4}
=
T_{p,\max}
\left(
1-qF_\star A_g
\right)^{1/4},
\end{equation}
defines the maximum allowed temperature for a given phase integral and geometric albedo. Note that we assume the maximum temperature cases, corresponding to no day-night heat redistribution and therefore, $f=2/3$ (see Equation \ref{Tmax_def}).

Surface reflectance laboratory measurements for several common silicate compositions predict low geometric albedos $A_g<0.1$ \citep{Essack2020}. However, magma-ocean surfaces are not necessarily dark: oxidized or FeO-rich compositions and roughness can produce higher reflectivities \citep{ModirroustaGalian2021}. We therefore treat $A_g = 0.1$ as an illustrative low-albedo case rather than as a firm assumption. If $A_g=0.1$ in each bandpass, reflected light would contribute approximately \reflAgone\ to both eclipse depths. After subtracting this contribution, the remaining thermal emission corresponds to a brightness temperature of $T_{b,\rm joint}(A_g=0.1) = \tbjointAgone$.

\subsection{Comparison to other lava worlds}

\begin{figure}[b]
    \centering
    \includegraphics[width=1\linewidth]{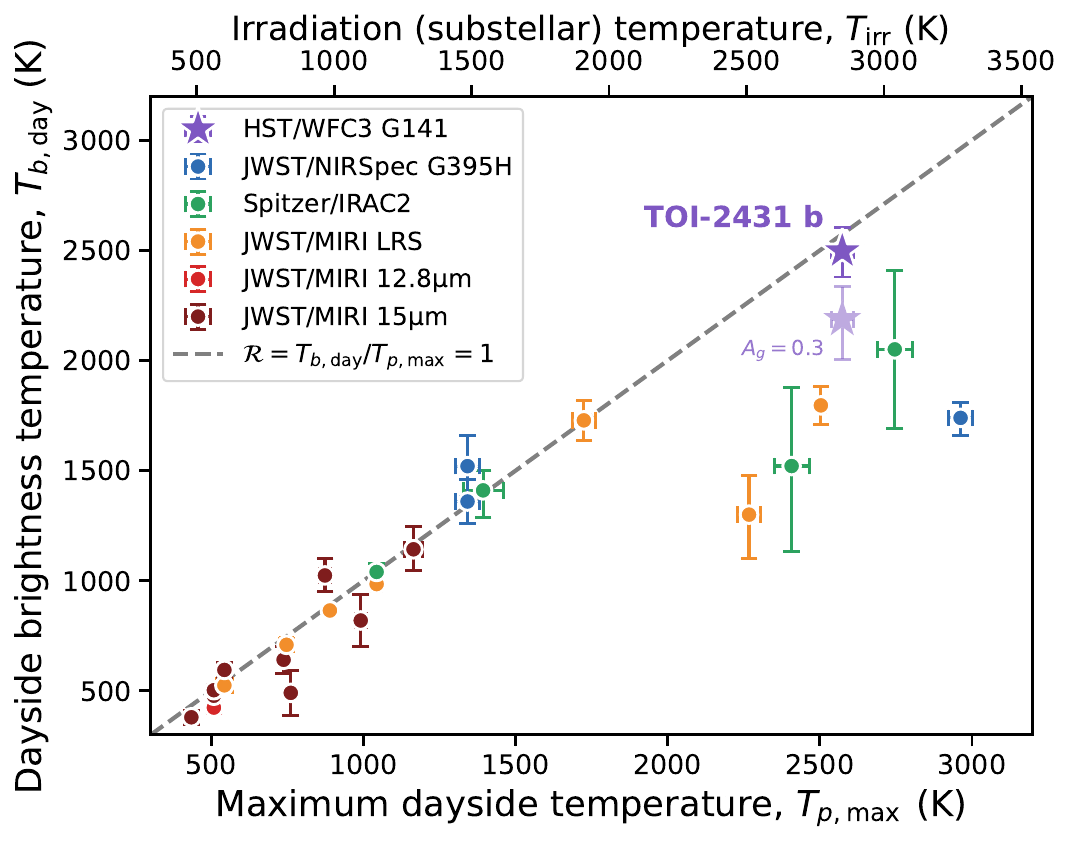}
    \caption{Published infrared dayside brightness temperatures, \(T_b\), of rocky exoplanets as a function of their zero Bond albedo, no-redistribution maximum dayside temperatures, \(T_{p,\max}\). The corresponding irradiation temperature, \(T_{\rm irr}\), is shown on the upper horizontal axis. The diagonal line marks \(\mathcal{R}=T_b/T_{p,\max}=1\), for which the measured brightness temperature equals the zero-albedo, no-redistribution limit. Error bars include the reported brightness-temperature uncertainties and the propagated uncertainties in \(T_\star\) and \(a/R_s\). Colors and symbols identify the observing instrument, as given in the legend. TOI-2431 b is highlighted and is shown under the assumption that its measured HST dayside flux is entirely thermal. A second semi-transparent data point for TOI-2431 b is shown below, assuming an arbitrary geometric albedo of $A_g = 0.3$. In order of increasing \(T_{\rm irr}\), the compilation includes TRAPPIST-1 c \citep{Zieba2023}, TRAPPIST-1 b \citep{Greene2023,Ducrot2025}, LTT 1445 A b \citep{Wachiraphan2025}, LHS 1140 c \citep{Rochon2026}, GJ 3929 b \citep{Connors2026}, GJ 1132 b \citep{Xue2024}, LHS 1478 b \citep{August2025}, TOI-1468 b \citep{MeierValdes2025}, GJ 486 b \citep{WeinerMansfield2024}, GJ 3473 b \citep{Holmberg2026}, LHS 3844 b \citep{Kreidberg2019,Zieba2026}, LTT 3780 b \citep{Allen2025}, TOI-1685 b \citep{Luque2025}, GJ 1252 b \citep{Crossfield2022}, GJ 367 b \citep{Zhang2024}, HD 3167 b \citep{Coy2026}, TOI-431 b \citep{Monaghan2025}, 55 Cnc e \citep{Hu2024}, TOI-2431 b (this work), K2-141 b \citep{Zieba2022}, and TOI-561 b \citep{Teske2025}.}
    \label{fig:Tmax}
\end{figure}

In Figure \ref{fig:Tmax}, we compare TOI-2431 b with all small exoplanets with published infrared secondary eclipse measurements, ranging from the temperate planet TRAPPIST-1 c \citep[$T_{\rm irr} = 480$ K;][]{Zieba2023} to TOI-561 b \citep[$T_{\rm irr} = 3300$ K;][]{Teske2025}. As first noted by \cite{Crossfield2022}, planets with temperatures below the threshold at which the substellar surface is expected to become fully molten generally have $\mathcal{R}\approx1$, consistent with low albedos and inefficient heat redistribution, although relatively thin (P$<$10 bar) secondary atmospheres may also be permitted \citep{Coy2026, Kreidberg2025}. At higher temperatures, several lava worlds appear to deviate from this trend. Their lower brightness temperatures may result from atmospheric heat redistribution, reflective surfaces or clouds, heat transport through the interior \citep{Meier2021, Meier2023, Herath2024, Farhat2025}, or a combination of these effects. 
Additional multiwavelength observations are needed to determine whether this apparent difference represents a physical transition \citep[e.g.,][]{WeinerMansfield2024}.

TOI-2431 b is not an obvious outlier in planet and host-star properties relative to the other highly irradiated planets in Figure \ref{fig:Tmax}. However, in particular, its broad mass and density posterior prevents comparison of these important parameters with those of the other planets, motivating further mass measurements for the planet.

\subsection{Pseudo-2D rock vapor model}\label{pseudo2Dmodeling}

We construct a pseudo-2D rock vapor model by splitting the planet into rings at different angles relative to the substellar point. Similar to \citet{Zieba2022}, we explore stellar zenith angles ranging from $0\deg$ (substellar point) to $80 \deg$, where a $90 \deg$ angle corresponds to the terminator region and assume zero incident flux there. We describe the approach in Section \ref{app:sec3_2D} of the Appendix.

At each of these stellar zenith angles, we self-consistently calculate the atmospheric surface pressure induced by rock vaporization and the surface temperature, using an iterative approach between the climate model {\texttt{ARCiS}} \citep{Ormel2019_ARCiS} and the vaporization tool {\texttt{LavAtmos}} \citep{vanBuchem2023}. 
The magma ocean is hereby modeled as an infinite reservoir of molten rock composed of a Bulk Silicate Earth (BSE) composition taken from \citet{2003TrGeo...2....1P}. This choice is motivated by the fact that, although we know little about the crust compositions of rocky exoplanets \citep{Miguel2011,vanBuchem2023}, the BSE provides a well-characterized Earth-like reference composition. We note, however, that different choices for melt composition can lead to different atmospheric compositions and temperature structures \citep{2024A&A...691A.159S,vanBuchem2026}. Elements included in our models are hence: O, Si, Ti, Fe, Ca, Al, K, Na, and Mg. We assume that no secondary volatile atmosphere is present on the planet, motivated by the high dayside temperature, pointing to weak heat redistribution.

The elemental abundances and surface pressure from vaporization serve as input parameters to the atmospheric modeling tool {\texttt{ARCiS}}, which runs radiative transfer and chemistry to generate temperature structures, emission spectra, and volume mixing ratios of molecular and atomic species. The star-planet system is modeled with parameters as listed in \citet{Tas2026}. We obtain the stellar spectrum using \texttt{PHOENIX} models and utilize gas phase opacities from the ExoMol database for relevant vapor species derived from  the BSE composition \citep{Chubb2021}. These species are: Na, SiO, SiO$_2$, TiO, Fe, K, O$_2$, O$_3$, O, MgO. For Fe, Mg, Si, and Ti, we take the opacities from Janssen et al. (submitted) computed with {\texttt{pyROX}} \citep{DeRegt2025}. All sources and their opacities are summarized in appendix \ref{opac_tables}.
We assume a zero surface albedo for TOI-2431 b as well as no additional heat flux from the interior of the planet.  The chemistry in the atmosphere is pure equilibrium chemistry and computed with {\texttt{GGchem}} \citep{Woitke2018_Equilibrium} iteratively with the temperature structure within {\texttt{ARCiS}}.

With this approach, we find that in general, the temperature of the atmosphere and surface decreases with zenith angle, see Figure \ref{Fig:partial_pressures}. As a consequence of this, the outgassing efficiency decreases as well, leading to thinner atmospheres further away from the substellar point. The maximum surface pressure is approximately 20\,mbar at the substellar point. This represents an upper limit on the equilibrium rock-vapor pressure across the modeled dayside under the adopted assumptions. The partial pressures of the most common species also decrease with increasing zenith angle as shown in Figure \ref{Fig:partial_pressures}. Na is the most abundant species throughout the modeled dayside, with O$_2$ and SiO also among the dominant species, and smaller abundances of TiO and Al. The models show atmospheric inversions throughout the dayside, consistent with strong absorption of stellar radiation by species such as Na, Fe, and TiO \citep{Zilinskas2022,vanBuchem2026}.

\subsection{Future JWST observations of TOI-2431 b} \label{sec:futureMIRI}

TOI-2431 b will be observed in JWST GO 8864 (PI: L. Dang) with a 10.6-hour full-orbit MIRI LRS phase curve that includes two secondary eclipses. 
These observations will extend the wavelength coverage to approximately 5--12 \micron, where the planetary flux is expected to be dominated by thermal emission. A joint MIRI, HST, and TESS analysis will therefore provide a substantially cleaner constraint on the dayside thermal emission and, indirectly, on the reflected-light contribution in the shorter-wavelength TESS and HST bandpasses.

In Figure \ref{fig:miri_spectrum}, we extend the blackbody models and our pseudo-2D rock-vapor atmosphere model to MIRI LRS wavelengths. The current TESS and HST measurements are not able to distinguish the rock-vapor spectrum from a blackbody. In the mid-infrared, the rock-vapor model predicts wavelength-dependent spectral structure in the MIRI LRS range. This is due to the opacities from species such as SiO and SiO$_2$, which are expected to produce features near 9 and 7 \micron, respectively \citep{Zilinskas2022,Zilinskas2023,vanBuchem2026}. MIRI LRS observations will therefore be able to measure the thermal contribution of the planet's emission and the presence of spectral features associated with a silicate atmosphere.

To estimate the improvement in the constraints on heat redistribution and albedo, we use \texttt{PandExo 2026.7} \citep{Batalha2017}. Combining the two planned eclipses as part of GO 8864, we predict a broadband eclipse depth of approximately \fpfsMIRIsim\ over 5--10.5 \micron\ for the $A_g=0$ and $T_{\rm day}=T_{p,\max}$ case. We use this simulated MIRI LRS measurement in Appendix \ref{app:miri_degeneracy} to quantify its expected impact on the inferred heat redistribution and albedos and present the respective posteriors in Figure \ref{fig:miri_degeneracy}.

\section{Conclusions}
We analyzed the HST/WFC3 G141 phase-curve observation of the ultra-short-period rocky planet TOI-2431 b and combined it with its TESS optical eclipse measurement. Our main conclusions are:

\begin{enumerate}
    \item We detect the secondary eclipse of TOI-2431 b with HST/WFC3, with an eclipse depth of \fpfsHSTobs\ over 1.127--1.570 \micron. This is the first secondary eclipse detection of a rocky exoplanet with HST and provides a near-infrared measurement of its dayside flux.
    
    \item Assuming planetary blackbody emission and negligible reflected light ($A_g = 0$), we find \(T_{b,\rm HST}=\tbHST\) and \(T_{b,\rm TESS}=\tbTESS\). A joint fit gives \(T_{b,\rm joint}=\tbjoint\) and \[ \mathcal{R}_{\rm joint} = \frac{T_{b,\rm joint}}{T_{p,\max}} = \rjoint, \] consistent with the zero-albedo, no-redistribution limit \(T_{p,\max}=\tpmax\).
    
    \item Under the same thermal-only, single temperature blackbody interpretation, the observations favor inefficient heat redistribution and low Bond albedo. We calculate one-sided \(2\sigma\) constraints of $f > \flower$ and $A_B < \ABupper$.

    \item The HST phase curve is also consistent with zero nightside emission and no measurable hotspot offset, with \(\phi=4.1^{+12.3}_{-12.6}\) degrees. These results are consistent with weak day-night redistribution.
    
    \item At the optical and near-infrared wavelengths covered by TESS and HST/WFC3, reflected light can play a significant part of the observed flux. Assuming a low geometric albedo of $A_g=0.1$ motivated by laboratory measurements \citep{Essack2020}, reflected light contributes approximately \reflAgone\ in each bandpass, and the joint brightness temperature decreases to \(T_{b,\rm joint}=\tbjointAgone\). More reflective solutions leading to lower dayside temperatures remain permitted by the current measurements.
    
    \item Under the thermal-dominated interpretation, TOI-2431 b clearly stands out from the emerging population of lava worlds with unexpectedly cool daysides. While several other observed lava worlds lie significantly below the zero albedo/no heat redistribution ($\mathcal{R}=1$) limit, TOI-2431 b's dayside temperature lies close to $T_{p, \rm{max}}$. 
    
    \item The forthcoming JWST/MIRI LRS phase curve \citep{Dang2025} will measure the planet at wavelengths where thermal emission should dominate. Combining MIRI with the HST and TESS measurements will substantially reduce the thermal--reflection degeneracy, improve constraints on heat redistribution and any hotspot offset, and search for departures from blackbody emission associated with a silicate atmosphere.
\end{enumerate}

TOI-2431 b therefore provides an important benchmark for interpreting the surfaces and atmospheres of highly irradiated rocky planets, which will be further characterized by the planned JWST/MIRI phase curve.

\begin{acknowledgments}
We thank the anonymous referee for their helpful comments. This paper includes data collected with the TESS mission, obtained from the MAST data archive at the Space Telescope Science Institute (STScI). Funding for the TESS mission is provided by the NASA Explorer Program. STScI is operated by the Association of Universities for Research in Astronomy, Inc., under NASA contract NAS 5-26555.
Based on observations made with the NASA/ESA Hubble Space Telescope, obtained from the Data Archive at the Space Telescope Science Institute, which is operated by the Association of Universities for Research in Astronomy, Inc., under NASA contract NAS5-26555. These observations are associated with program \#16660.
This research has made use of the NASA Exoplanet Archive, which is operated by the California Institute of Technology, under contract with the National Aeronautics and Space Administration under the Exoplanet Exploration Program.
Some of the computations in this paper were conducted on the Smithsonian High Performance Cluster (SI/HPC), Smithsonian Institution \url{https://doi.org/10.25572/SIHPC}. All of the data presented in this article were obtained from the Mikulski Archive for Space Telescopes (MAST) at the Space Telescope Science Institute. The specific observations analyzed can be accessed via \dataset[doi:10.17909/fad2-yf80]{https://doi.org/10.17909/fad2-yf80}.\\
In addition, this work used the Dutch national e-infrastructure with the support of the SURF Cooperative using grant no. EINF-15133.
S.Z. was supported by NASA through the NASA Hubble Fellowship grant \#HST-HF2-51570.001-A awarded by the Space Telescope Science Institute, which is operated by the Association of Universities for Research in Astronomy, Incorporated, under NASA contract NAS5-26555. 
L.D. acknowledges support from the Natural Sciences and Engineering Research Council (NSERC), the Canadian Space Agency, and the Waterloo Centre for Astrophysics.
M. L-M. is supported by individual research time under NASA contracts NAS5-26555 and NAS5-03127 to the Associated Universities for Research in Astronomy for the operation of the Hubble Space Telescope and James Webb Space Telescope Science Operations Centers at STScI.
O.H. acknowledges financial support by FFG under grant number 62545255.
We used ChatGPT (GPT-5.6 Sol, OpenAI) under author supervision to assist with plotting code and figure refinement. All resulting code was reviewed and verified by the authors.
\end{acknowledgments}

\facilities{HST(WFC3), TESS}

\software{
\texttt{PACMAN} \citep{Zieba2022_PACMAN}, \texttt{pysynphot} \citep{pysynphot2013}, \texttt{stsynphot} \citep{stsynphot2020}, \texttt{dynesty} \citep{Speagle2020, Koposov2025}, \texttt{batman} \citep{Kreidberg2015}, \texttt{speclib} \citep{Rackham2024, Rackham2026}, \texttt{PandExo} \citep{Batalha2017}, \texttt{Iraclis} \citep{Tsiaras2016, Tsiaras2018}, \texttt{ExoTETHyS} \citep{morello_exotethys}, \texttt{pymultinest} \citep{feroz_multinest,buchner_multinest}, \texttt{celerite2} \citep{ForemanMackey2017}, \texttt{emcee} \citep{ForemanMackey2013}
}

\appendix

\section{\texttt{PACMAN} HST reduction}
\label{app:sec1}

\subsection{\texttt{PACMAN} reduction choices}
\label{app:sec1_reduction}

Here, we report the \texttt{PACMAN} tests that were performed to select the final configuration, which we adopted in Section \ref{sec:pacman}, by exploring various background subtractions and apertures. We selected the setup that minimized the RMS of the final best-fit residuals. For the background subtraction, we compared several local background boxes with an estimate based on the median flux of pixels outside the region containing the scanned spectrum. The latter resulted in slightly lower residual RMS and was therefore adopted.
For the optimal extraction, we tested different aperture sizes and adopted an aperture extending 21 pixels beyond the upper and lower edges of the spectrum, which minimized the residual RMS.
We note that the smoothed spectrum for the wavelength calibration is only used to determine the wavelength shift across the G141 bandpass, and that the calibration is expected to be insensitive to differences in individual stellar absorption lines.
Because of an observing setup issue identified after the observations, part of the spectral trace fell outside the detector. We therefore tested different wavelength ranges and adopted 1.127--1.570 \micron, which provided the lowest residual RMS while avoiding the detector edge.

\subsection{\texttt{PACMAN} treatment of systematics}
\label{app:sec1_1}

It has been shown that the choice of the visit-long systematic trend can impact the recovered transit or eclipse depth \citep[e.g.,][]{Stevenson2014, Guo2020, Edwards2024}. Hence, we tested linear, quadratic, exponential, and logarithmic trends:

\begin{equation} \label{eqn:fitting_model_visit}
F_\textrm{sys,v}(t) =
\begin{cases}
C(t) (1 + v_1 t_\textrm{v}),&\rm linear \\
C(t) (1 + v_1 t_\textrm{v} + v_2 t_\textrm{v}^2),&\rm quadratic \\ 
C(t) (1 - e_1 \exp(-  \, t_\textrm{v}/e_2)),&\rm exponential \\ 
C(t) (1 - l_1 \log(t_\textrm{v} + l_2)),&\rm logarithmic. \\

\end{cases}
\end{equation}

Here, $v_1$, $v_2$, $e_1$, $e_2$, $l_1$, and $l_2$ are the free parameters of the corresponding visit-long trends, and \(t_{\mathrm{v}}\) is the time elapsed since the first integration of the visit. The constant \(C(t)\) depends on the scan direction. For forward scans \(C(t) = c\), while for reverse scans, \(C(t) = c (1 + \mathrm{scale})\).

For the orbit-long trend, we use 

\begin{equation} \label{eqn:fitting_model_orbit}
    F_\textrm{sys,o}(t) = 1 - \exp(-r_1 t_\textrm{o} - r_2 - D(t)),
\end{equation}

where $r_1$ and $r_2$ describe the ramp for every HST orbit. The term $D(t)$ accounts for the stronger ramp often observed in the first fitted orbit. $D(t)$ has the value $r_3$ in this first fitted orbit and is zero for all subsequent orbits. The time that elapsed since the first fitted integration in an orbit is given by $t_\textrm{o}$. 

\subsection{\texttt{PACMAN} model selection}
\label{app:sec1_2}

For all fits we use \texttt{dynesty} and set \texttt{nlive\_init} = 2000, \texttt{dlogz\_init} = 0.0001, and \texttt{nlive\_batch} = 2000, corresponding to the initial number of live points, the initial stopping criterion, and the number of live points used in each subsequent dynamic batch, respectively.

Our model comparison yields several initial results from the data. We define
\(\Delta\log\mathcal{Z}=\log\mathcal{Z}_{\mathrm{model\_baseline}}-\log\mathcal{Z}_{\mathrm{model\_new}}\), such that positive values favor the baseline model. Following the empirical scale presented by \citet{Trotta2008}, we interpret $\Delta\log\mathcal{Z}\geq2.5$ as moderate evidence and $\Delta\log\mathcal{Z}\geq5$ as strong evidence. These thresholds, however, only provide qualitative guidelines rather than strict detection criteria \citep{Thorngren2026}. We note that other than the comparison between the different visit-long trends themselves, we perform the following model selection tests using the adopted exponential visit-long trend model.

\begin{itemize}

\item We find strong evidence for the presence of the combined eclipse and phase-curve signal. Comparing a model containing a \texttt{batman} eclipse and a cosine phase curve model, to one without either component, we find \(\Delta\log\mathcal{Z}=9.7\), therefore finding strong evidence in favor of the model containing the planetary emission signal. Also when using a model that only has the phase variation but no eclipse, we find \(\Delta\log\mathcal{Z}=11.2\), therefore finding strong evidence in favor of the eclipse again.

\item We find no evidence for nightside emission or a phase-curve offset. The model with \(\mathrm{AmpCos}=1\) and \(\mathrm{AmpSin}=0\) is preferred over a model in which only \(\mathrm{AmpCos}\) is free by \(\Delta\log\mathcal{Z}=\mathrm{3.1}\), constituting moderate evidence, and over a model in which both \(\mathrm{AmpCos}\) and \(\mathrm{AmpSin}\) are free by \(\Delta\log\mathcal{Z}=\mathrm{8.1}\), constituting strong evidence. When both coefficients are fitted, we find \(\mathrm{Amp}=\mathrm{0.88^{+0.24}_{-0.19}}\) and \(\phi=\mathrm{4.1{^{\circ}}^{+12.3{^{\circ}}}_{-12.6{^{\circ}}}}\), consistent with no nightside emission (\(\mathrm{Amp}=1\)) and no phase offset (\(\phi=\ 0^{\circ}\)).

\item We test if the inferred phase curve amplitude depends on the adopted visit-long trend. For this we fit for AmpCos and then for the exponential, logarithmic, quadratic, and linear visit-long models. All obtained values for AmpCos are consistent with AmpCos = 1. We then test if the preferred exponential model parameters ($e_1$ and $e_2$) are correlated with AmpCos. We find Pearson correlation coefficients of 0.02 and -0.08, where +1 would imply perfect positive correlation and -1, perfect anticorrelation. Our results suggest no noticeable correlation between AmpCos and the exponential trend. We also check the preference between having a fixed or free AmpCos parameter. The fixed AmpCos model (AmpCos$ = 1$) is preferred by \(\Delta\log\mathcal{Z}=\mathrm{3.1, 1.4, }\) and $1.8$ for the exponential, quadratic and linear trends respectively. The logarithmic model weakly (and not significantly) favors the free AmpCos model by \(\Delta\log\mathcal{Z}=\mathrm{1.0}\). Finally, we find that the inferred eclipse depths for all visit-long trends are consistent between the fixed- and free-AmpCos fits.

\item The first orbit exhibits a stronger ramp, represented by the \(D(t)\) parameter in Equation \ref{eqn:fitting_model_orbit}. The model including this term is preferred over one with \(r_3=0\) by \(\Delta\log\mathcal{Z}=24.0\), providing strong evidence for the additional first-orbit term.

\item A linear limb-darkening law is only weakly preferred over a quadratic law, parameterized following \citet{Kipping2013}, with \(\Delta\log\mathcal{Z}=\mathrm{1.9}\). The evidence difference is inconclusive; we therefore adopt the simpler linear limb-darkening law. Note that the transit depths resulting from the linear and quadratic laws are consistent with each other. Because the transit depth is not used in our following scientific discussion, the choice of limb-darkening law does not affect our conclusions.

\item A circular orbit is moderately preferred over an eccentric orbit, with \(\Delta\log\mathcal{Z}=\mathrm{3.0}\). When eccentricity is allowed to vary, it remains poorly constrained, with a \(3\sigma\) upper limit of \(e<0.19\). A circular orbit is also consistent with the short tidal circularization timescale expected for an ultra-short-period planet \citep{Winn2018}. We therefore set the eccentricity to zero for all following fits.

\item The exponential visit-long trend is preferred over the linear, quadratic, and logarithmic trends by \(\Delta\log\mathcal{Z}=\mathrm{17.1}\), \(\mathrm{8.1}\), and \(\mathrm{6.8}\), respectively. All three provide strong evidence for the exponential trend, which we therefore adopt for the remaining fits.

\end{itemize}

We also considered marginalizing the eclipse depth over the tested systematics models, following the approach of \citet{Wakeford2016}. However, the exponential model carries 99.9\% of the total evidence weight compared to the other systematic models, leading to a negligible change in eclipse depth from that of the adopted exponential model.

\section{\texttt{Iraclis} HST reduction}
\label{app:sec2}

\begin{figure}[h]
    \centering
    \includegraphics[width=0.87\linewidth]{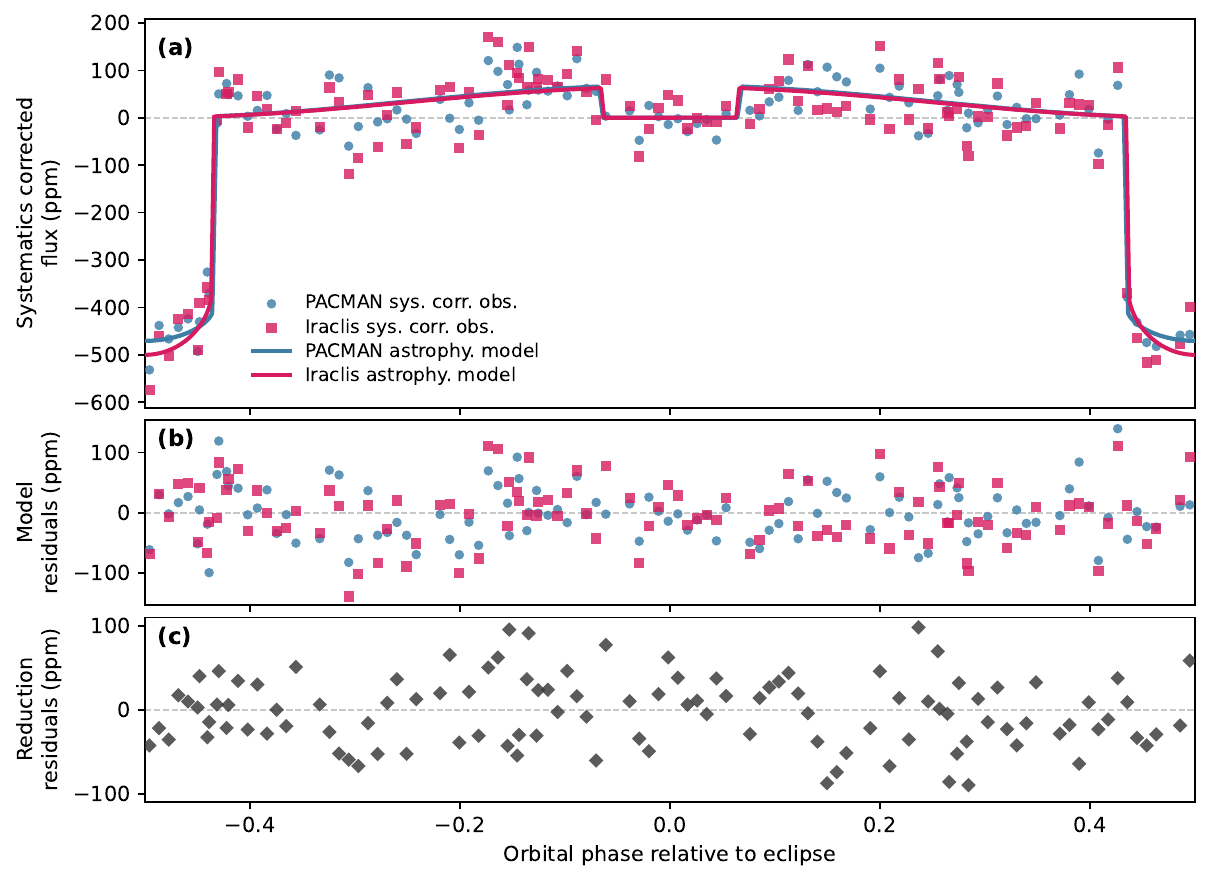}
    \caption{\texttt{Iraclis} reduction of the HST/WFC3 G141 white-light phase curve of TOI-2431\,b compared to the \texttt{PACMAN} reduction. (a) Systematics-corrected light curves and best-fit astrophysical models for the two reductions, with \texttt{PACMAN} shown as blue circles and \texttt{Iraclis} as red squares. Differences between the models during transit are due to their different treatments of limb darkening. The transit depth is, however, not used in our phase-curve analysis. (b) Residuals after subtracting the respective best-fit astrophysical models from the systematics-corrected light curves. (c) Difference between the systematics-corrected \texttt{Iraclis} and \texttt{PACMAN} light curves.}
    \label{fig:irac}
\end{figure}

We also performed an independent reduction with \texttt{Iraclis}. We performed two fits to the extracted light curve, one assuming AmpCos = 1 and AmpSin = 0 as well as a second where these were both free parameters. From the latter fit, we recover AmpCos = 0.86$^{+0.10}_{-0.15}$
and AmpSin = -0.20$^{+0.23}_{-0.27}$ 
(Amp = 0.91$^{+0.10}_{-0.14}$, 
$\phi$ = $-13{^{\circ}}^{+15{^{\circ}}}_{-17{^{\circ}}}$)
, a result that is consistent with no nightside emission (Amp=1) and no hotspot offset ($\phi = 0{^{\circ}}$). The model where AmpCos and AmpSin are fixed is preferred by \(\Delta\log\mathcal{Z}=\mathrm{1.66}\), indicating only a weak preference. Both models give similar eclipse depths (60$\pm$14 ppm when AmpCos/AmpSin are fixed, 59$\pm$17 ppm when they are free), indicating that the choice of model does not impact the inferences on the dayside temperature. In Figure \ref{fig:irac}, we show the best-fit model (AmpCos = 1 and AmpSin = 0) to the independent \texttt{Iraclis} reduction and how it compares to the \texttt{PACMAN} reduction.

\section{TESS Reduction}
\begin{figure}[h]
    \centering
    \includegraphics[width=0.66\linewidth]{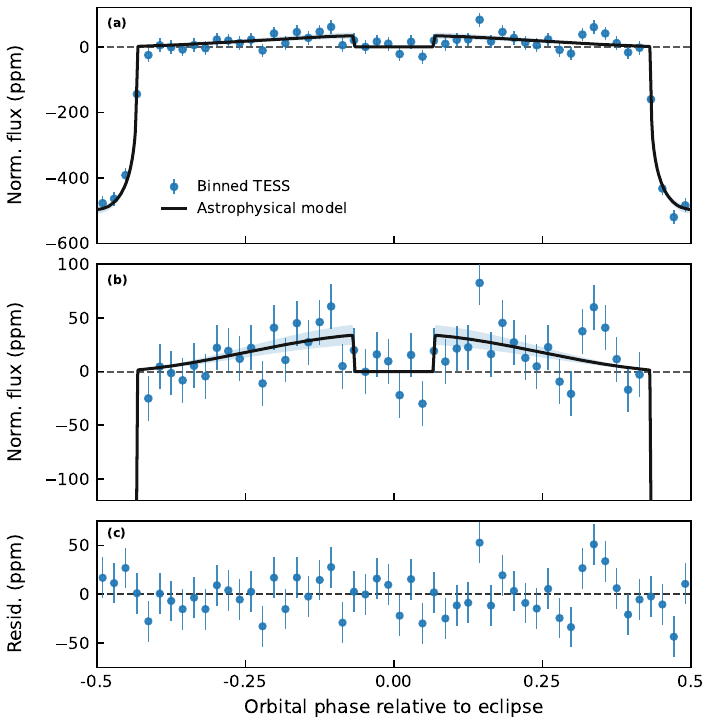}
    \caption{TESS phase curve of TOI-2431 b. (a) The GP-corrected phase-folded light curve. (b) Zoom on the eclipse and phase variation. (c) The residuals from the best-fitting model.}
    \label{fig:TESS}
\end{figure}

\section{Opacities and sources}
\label{opac_tables}

\begin{table}[h]
  \centering
  \caption{Opacity sources used in the pseudo-2D rock-vapor model}
  \begin{tabular}{ll}
    \hline
    Species & Reference \\% & link \\
    \hline
    \texttt{O3} &     {\cite{2022JQSRT.27707949G}} \\% & \href{https://ui.adsabs.harvard.edu/abs/2022JQSRT.27707949G/abstract}{ADS link} \\
    \texttt{O2} &     {\cite{2022JQSRT.27707949G}} \\% & \href{https://ui.adsabs.harvard.edu/abs/2022JQSRT.27707949G/abstract}{ADS link} \\
    \texttt{O} &     {\cite{Kuruzc2018}} \\% & \href{https://ui.adsabs.harvard.edu/abs/2018ASPC..515...47K}{ADS link} \\
    \texttt{SiO} &     {\cite{2022MNRAS.510..903Y}} \\% & \href{https://ui.adsabs.harvard.edu/abs/2022MNRAS.510..903Y/abstract}{ADS link} \\
    \texttt{Na} &     {\cite{Allard2019_New}} \\% & \href{https://ui.adsabs.harvard.edu/abs/2019A%26A...628A.120A/abstract}{ADS link} \\
    \texttt{K} &     {\cite{Allard2016_KH2}} \\% & \href{https://ui.adsabs.harvard.edu/abs/2016A%26A...589A..21A/abstract}{ADS link} \\
    \texttt{TiO} &     {\cite{2019MNRAS.488.2836M}} \\% & \href{https://ui.adsabs.harvard.edu/abs/2019MNRAS.488.2836M/abstract}{ADS link} \\
    \texttt{MgO} &     {\cite{2019MNRAS.486.2351L}} \\% & \href{https://ui.adsabs.harvard.edu/abs/2019MNRAS.486.2351L/abstract}{ADS link} \\
    \texttt{SiO2} &     {\cite{2020MNRAS.495.1927O}} \\% & \href{https://ui.adsabs.harvard.edu/abs/2020MNRAS.495.1927O/abstract}{ADS link} \\
    \texttt{Mg} &     {\cite{Kuruzc2018}} \\% & \href{https://ui.adsabs.harvard.edu/abs/2018ASPC..515...47K}{ADS link} \\
    \texttt{Fe} &     {\cite{Kuruzc2018}} \\% & \href{https://ui.adsabs.harvard.edu/abs/2018ASPC..515...47K}{ADS link} \\
    \texttt{Al} &     {\cite{Kuruzc2018}} \\% & \href{https://ui.adsabs.harvard.edu/abs/2018ASPC..515...47K}{ADS link} \\
    \texttt{Ti} &     {\cite{Kuruzc2018}} \\% & \href{https://ui.adsabs.harvard.edu/abs/2018ASPC..515...47K}{ADS link} \\
    \texttt{Si} &     {\cite{Kuruzc2018}} \\% & \href{https://ui.adsabs.harvard.edu/abs/2018ASPC..515...47K}{ADS link} \\
    \hline
    \end{tabular}
    \label{Tab:opacities}
\end{table}

\begin{table}[h]
  \centering
  \caption{Modeling tools used for the pseudo-2D rock-vapor model}
  \begin{tabular}{ll}
    \hline
    Code & Reference \\% & link \\
    \hline
    \texttt{ARCiS} &     {\cite{Min2020_ARCiS}} \\% & \href{https://ui.adsabs.harvard.edu/abs/2020A%26A...642A..28M/abstract}{ADS link} \\
    \texttt{GGchem} &     {\cite{Woitke2018_Equilibrium}} \\% & \href{https://ui.adsabs.harvard.edu/abs/2018A%26A...614A...1W/abstract}{ADS link} \\
    \texttt{ExoMolOP} &     {\cite{Chubb2021_ExoMolOP}} \\% & \href{https://ui.adsabs.harvard.edu/abs/2021A&A...646A..21C}{ADS link} \\
    \texttt{pyROX} &     {\cite{DeRegt2025}} \\% & \href{https://ui.adsabs.harvard.edu/abs/2025arXiv251020870D}{ADS link} \\
    \hline
    \end{tabular}
    \label{Tab:references_ARCiS}
\end{table}

\section{Construction of Pseudo-2D model}
\label{app:sec3_2D}

To construct a disk-integrated dayside emission spectrum from the models run at various angles relative to the substellar spot (i.e., zenith angles), we divide the visible hemisphere into concentric rings centered on the substellar point, similarly to \citep{Zieba2022}. We first calculate local planet-to-star flux-ratio spectra at stellar zenith angles $\theta=0^\circ,10^\circ,\ldots,80^\circ$. We assume that the stellar flux vanishes at the terminator, i.e., $(f_p/f_s)_{90^\circ}=0$, neglecting hyperillumination and horizontal heat redistribution. To obtain a representative spectrum for each $10^\circ$-wide ring, we linearly interpolate the neighboring model spectra in the irradiation coordinate $\beta=\cos\theta$. The resulting spectra at $\theta=5^\circ,15^\circ,\ldots,85^\circ$ represent the rings spanning $0^\circ$--$10^\circ$, $10^\circ$--$20^\circ$, ..., $80^\circ$--$90^\circ$, respectively.

The projected area of an infinitesimal annulus at zenith angle $\theta$ is
\begin{equation}
{\rm d}A_{\rm proj}
=
2\pi R_p^2\sin\theta\cos\theta,{\rm d}\theta .
\end{equation}

Normalizing the projected area of each ring by the projected area of the planetary disk, $\pi R_p^2$, gives the weight
\begin{equation}
w_i
=
\frac{1}{\pi R_p^2}
\displaystyle\int_{\theta_{i,1}}^{\theta_{i,2}}
2\pi R_p^2\sin\theta\cos\theta,{\rm d}\theta
=
\sin^2\theta_{i,2}-\sin^2\theta_{i,1}.
\end{equation}

The disk-integrated eclipse spectrum is then calculated wavelength by wavelength as

\begin{equation}
\left(\frac{f_p}{f_s}\right)_{\rm eclipse}(\lambda)
=
\sum_i
w_i
\left(\frac{f_p}{f_s}\right)_i(\lambda),
\end{equation}
where $\sum_i w_i=1$.

\section{TP Profiles and Species Abundances}
\begin{figure*}
    \centering
    \includegraphics[width=1\linewidth]{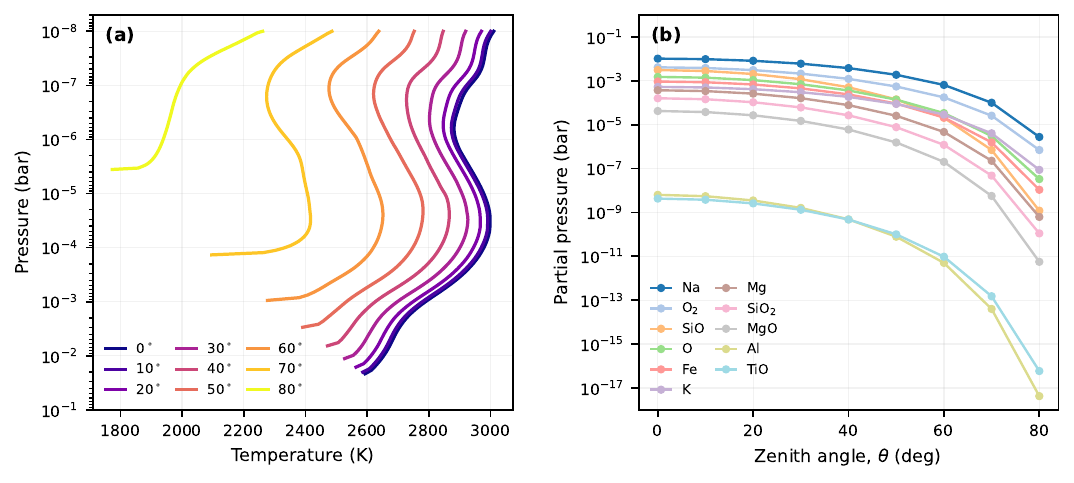}
    \caption{Results of the pseudo-2D rock-vapor modeling for TOI-2431 b. (a) Temperature-pressure profiles obtained from the self-consistent pseudo-2D models described in Section \ref{pseudo2Dmodeling}. We assume zero surface albedo, no additional internal heat flux, and no horizontal heat redistribution. A stellar zenith angle of $0\deg$ corresponds to the substellar point, while $90\deg$ corresponds to the terminator. As rock vaporization becomes less efficient with increasing zenith angle, the surface pressures decrease towards the terminator. (b) Surface partial pressures of selected atmospheric species for a BSE composition as a function of stellar zenith angle. These partial pressures are calculated with \texttt{LavAtmos} \citep{vanBuchem2023}, using the surface temperatures obtained from the self-consistent pseudo-2D models.}
    \label{Fig:partial_pressures}
\end{figure*}

\section{JWST MIRI/LRS breaking the degeneracy}
\label{app:miri_degeneracy}

\begin{figure*}[h]
    \centering
    \includegraphics[width=1\linewidth]{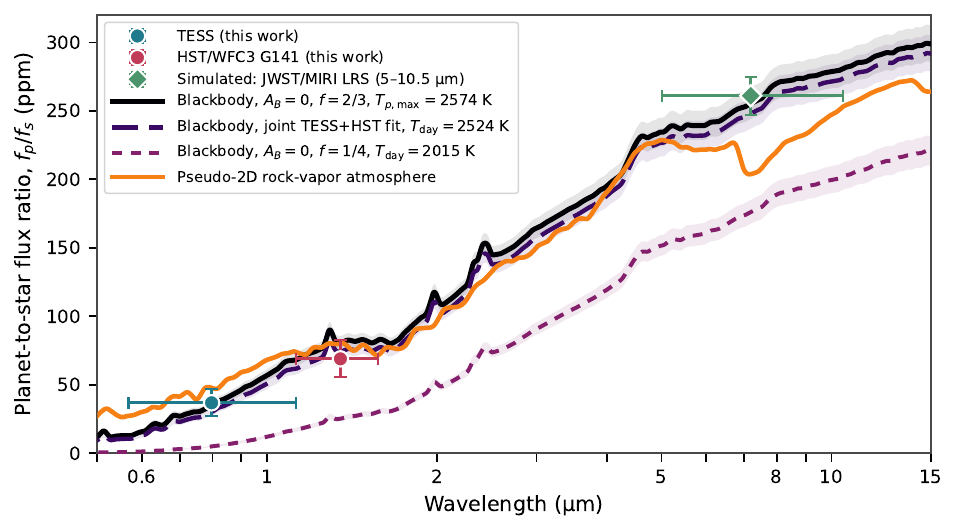}
    \caption{Predicted emission spectrum of TOI-2431 b from the optical to the JWST/MIRI LRS wavelength range. The TESS and HST measurements and the three blackbody models are the same as in Figure \ref{fig:spectrum}, but are extended here to the mid-infrared. The green diamond shows a simulated MIRI LRS broadband eclipse measurement assuming $A_g=0$ and $T_{\rm day}=T_{p,\max}$ and limiting the analysis to the 5--10.5 \micron\ interval leading to \fpfsMIRIsim. Unlike the single broadband point shown here, MIRI LRS provides spectroscopic information across this wavelength range giving the opportunity to search for features caused by a rock vapor atmosphere.}
    \label{fig:miri_spectrum}
\end{figure*}

\begin{figure*}[h]
    \centering
    \includegraphics[width=1\linewidth]{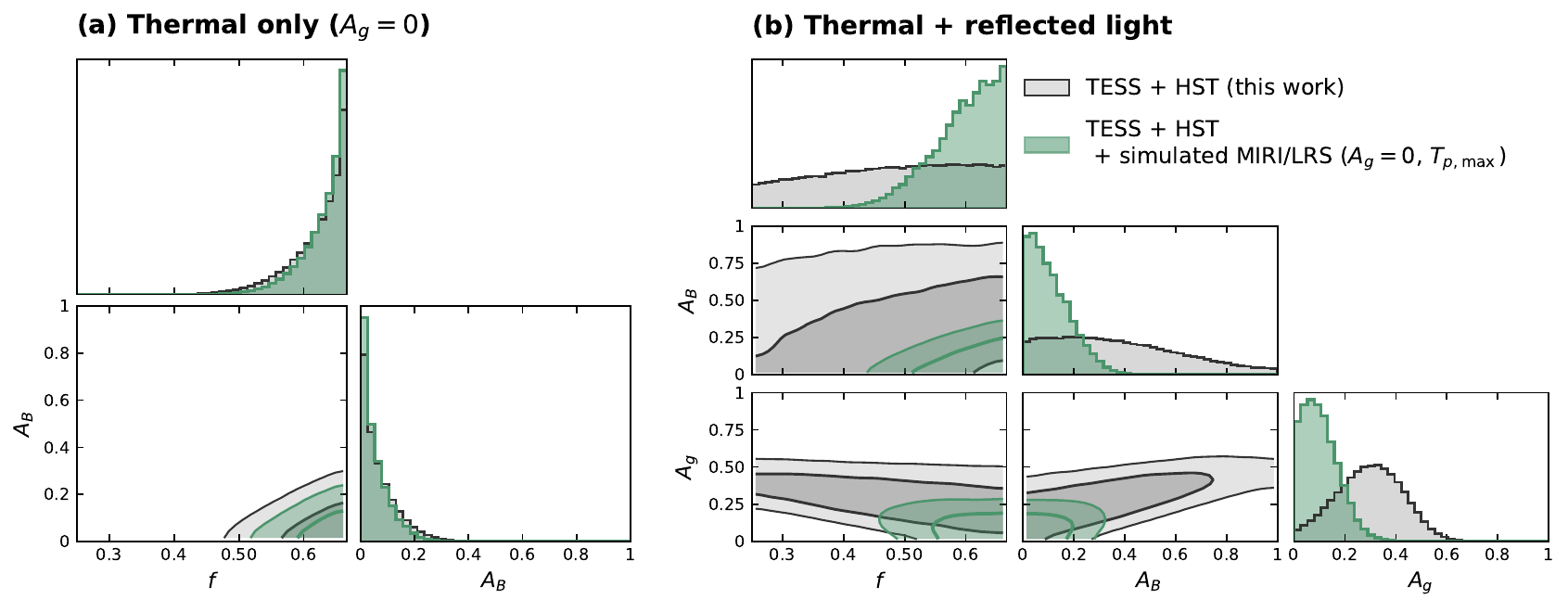}
    \caption{Expected improvement in the energy-budget constraints from the planned JWST/MIRI LRS observations. Gray distributions show the current TESS+HST posteriors, while green distributions include the simulated 5--10.5 \micron\ MIRI LRS eclipse measurement shown in Figure \ref{fig:miri_spectrum}, which assumes $A_g=0$ and $T_{\rm day}=T_{p,\max}$. Panel (a) assumes thermal emission only and fits $f$ and $A_B$. Panel (b) additionally includes reflected light and independently fits $f$, $A_B$, and $A_g$, without imposing any relation between $A_B$ and $A_g$.}
    \label{fig:miri_degeneracy}
\end{figure*}

Next, we estimate the improvement in the energy-budget constraints from the two secondary eclipses contained in the planned JWST/MIRI LRS phase curve. As described in Section \ref{sec:futureMIRI}, we use \texttt{PandExo 2026.7} \citep{Batalha2017} to estimate a combined eclipse depth precision of \miriPrecision\ over 5--10.5 \micron. Using $A_g=0$ and $T_{\rm day}=T_{p,\max}$ leads to a band-integrated eclipse depth of approximately \fpfsMIRIsim.

Figure \ref{fig:miri_spectrum} compares this simulated measurement to the TESS and HST eclipse depths and the corresponding model emission spectra. At the current TESS and HST wavelengths, the rock-vapor model is difficult to distinguish from the blackbody solutions. At longer wavelengths, however, the model develops spectral features around 7--8 \micron\ due to SiO$_2$ \citep{Zilinskas2022}.

We analyze how this MIRI LRS measurement would affect the inferred energy budget. We consider two cases. First, we assume thermal emission only and fit for the heat-redistribution factor $f$ and Bond albedo $A_B$. Second, we additionally include reflected light and independently fit $f$, $A_B$, and the geometric albedo $A_g$. In the latter case, we do not impose a relation between $A_B$ and $A_g$. Figure \ref{fig:miri_degeneracy} compares the resulting posteriors with those obtained from the current TESS+HST measurements alone.

For the thermal-only case, adding the simulated MIRI LRS measurement slightly narrows the allowed ranges of both $f$ and $A_B$. When reflected light is included, the current TESS+HST measurements permit a considerably broader range of $f$, $A_B$, and $A_g$. Adding the simulated MIRI LRS measurement strongly reduces this parameter space and recovers the low-albedo, inefficient-redistribution region used to generate the simulated measurement. For example, the width of the $A_g$ posterior decreases by approximately a factor of two. This illustrates how the longer-wavelength MIRI LRS observations can break the thermal--reflection degeneracy present in the TESS and HST measurements.

\bibliography{references}{}
\bibliographystyle{aasjournalv7}

\end{document}